\documentclass[12pt,draftcls,journal,onecolumn]{IEEEtran}
\usepackage[T1]{fontenc}
\usepackage[latin9]{inputenc}
\usepackage{color}
\usepackage{amsmath}

\usepackage{graphicx}
\usepackage{epsfig}
\usepackage{amssymb}
\usepackage{amsthm}
\begin{document}

\title{Structured Dispersion Matrices from Space-Time Block Codes for Space-Time Shift Keying }
\author{
      \IEEEauthorblockN{Rakshith Rajashekar,~\IEEEmembership{Student Member,~IEEE}, K.V.S. Hari,~\IEEEmembership{Senior Member,~IEEE}, and L. Hanzo,~\IEEEmembership{Fellow,~IEEE}
        \thanks{Rakshith Rajashekar and K.V.S.Hari are with the Department
of Electrical Communication Engineering, Indian Institute of Science, India     
  (e-mail: \{ rakshithmr, hari\}@ece.iisc.ernet.in).}}

   \thanks{L. Hanzo is with the School of ECS, University of Southampton, UK
  (e-mail: lh@ecs.soton.ac.uk).\newline A part of this work was presented in the proceedings of {\em IEEE GLOBECOM conf.}, 2011.}
\thanks{ The financial support of the DST, India and of the EPSRC, UK under the auspices
 of the India-UK Advanced Technology Center (IU-ATC) is gratefully acknowledged.}}

\maketitle

\begin{abstract}
Coherent Space-Time Shift Keying (CSTSK) is a recently developed generalized shift-keying framework for Multiple-Input Multiple-Output systems, which uses a set of Space-Time matrices termed as Dispersion Matrices (DM). CSTSK may be combined with a classic signaling set (eg. QAM, PSK) in order to strike a flexible tradeoff between the achievable diversity and multiplexing gain. One of the key benefits of the CSTSK scheme is its Inter-Channel Interference (ICI) free system that makes single-stream Maximum Likelihood detection possible at low-complexity. In the existing CSTSK scheme, DMs are chosen by maximizing the mutual information over a large set of complex valued, Gaussian random matrices through numerical simulations. We refer to them as Capacity-Optimized (CO) DMs. In this contribution we establish a connection between the STSK scheme as well as the Space-Time Block Codes (STBC) and show that a class of STBCs termed as Decomposable Dispersion Codes (DDC) enjoy all the benefits that are specific to the STSK scheme. Two STBCs belonging to this class are proposed, a rate-one code from Field Extensions and a full-rate code from Cyclic Division Algebras, that offer structured DMs with desirable properties such as full-diversity, and a high coding gain. We show that the DMs derived from these codes are capable of achieving a performance than CO-DMs, and emphasize the importance of DMs having a higher coding gain than CO-DMs in scenarios having realistic, imperfect channel state information at the receiver. 
\end{abstract}

\begin{keywords}
Space-time block code, space-time shift keying, coding gain, diversity, STBCs from division algebras.
\end{keywords}

\section{Introduction}
 
\IEEEPARstart SPATIAL MODULATION (SM) \cite{SM1}, \cite{SM2} is a novel development in the family of low complexity Multiple-Input Multiple-Output (MIMO) schemes that exploits the MIMO channel for transmitting information in an unprecedented fashion. This scheme has attracted the attention of various researchers and led to a number of novel schemes, such as Space-Shift Keying (SSK) \cite{SSK1}, Coherent Space-Time Shift Keying (CSTSK) \cite{CSTSK}, \cite{GSTSK}, Time-Orthogonal Signal Design assisted Spatial Modulation (TOSD-SM) \cite{TOSD-SM}, \cite{TOSD-SSK} and Space-Time Block Coded Spatial Modulation (STBC-SM) \cite{STBC-SM}. The key benefits offered by the SM/SSK schemes are that no Inter-Antenna Synchronization (IAS) is required and the Inter-Channel Interference (ICI) is readily controllable at the receiver. Hence, low-complexity single-stream based Maximum Likelihood (ML) detection may be used \cite{LCML}. However, some of these schemes fail to offer a transmit diversity, since only a single transmit antenna is activated in any symbol duration. The CSTSK, TOSD-SM, and STBC-SM schemes were some of the schemes proposed for increasing the transmit diversity order beyond one in the family of SM/SSK schemes. The TOSD-SM scheme uses time-orthogonal shaping filters that attains a transmit diversity order of two. Higher diversity orders were shown to be possible, but at the cost of requiring IAS at the transmitter. The STBC-SM scheme uses an STBC like Alamouti code spreading the signal to the space; time; and spatial domain for achieving a transmit diversity order of more than one.

CSTSK is capable of striking a flexible tradeoff between the attainable diversity and multiplexing gain~\cite{CSTSK},~\cite{GSTSK}. This scheme was shown to exhibit a better performance than the SM and SSK schemes, since it is capable of achieving both transmit- and receive-diversity. Although, the CSTSK scheme potentially requires IAS, it enjoys the benefit of low-complexity ML detection due to its ICI-free system model. The information bits in this scheme are first partitioned into two sets, and then one of the sets is mapped to a point from a conventional signal set like $L$-QAM, or $L$-PSK, while the other set of bits to the index of a matrix from a set of $Q$ Dispersion Matrices (DM). Specifically, the CSTSK scheme activates one out of $Q$ $(M \times T)$-element DMs, which is then multiplied by one of the legitimate symbols from an $L$-symbol constellation, where $T$ is the number of time-slots.
This scheme offers a throughput independent of $M$, given by 
\begin{equation}
 R_{CSTSK}=\frac{\log_2{(Q\cdot L)}}{T} ~\text{bpcu},
\end{equation}
where bpcu in short for bits/channel use.
The DMs in the existing scheme \cite{CSTSK} are chosen by maximizing the mutual information over a large set of unity-average-power, complex valued, Gaussian random matrices. We refer to them as Capacity-Optimized DMs (CO-DM). Since, the designs generated this way for maximizing the capacity are unable to guarantee achieving the maximum attainable coding gain, they do not necessarily minimize the Symbol Error Rate (SER)~\cite{APaul}, \cite{HEATH}.  The focus of this paper is to design structured DMs that attain a better bit error ratio (BER) performance than that given by the DMs of the existing scheme.
\newline

\begin{figure}
\centering
\includegraphics[scale=0.3]{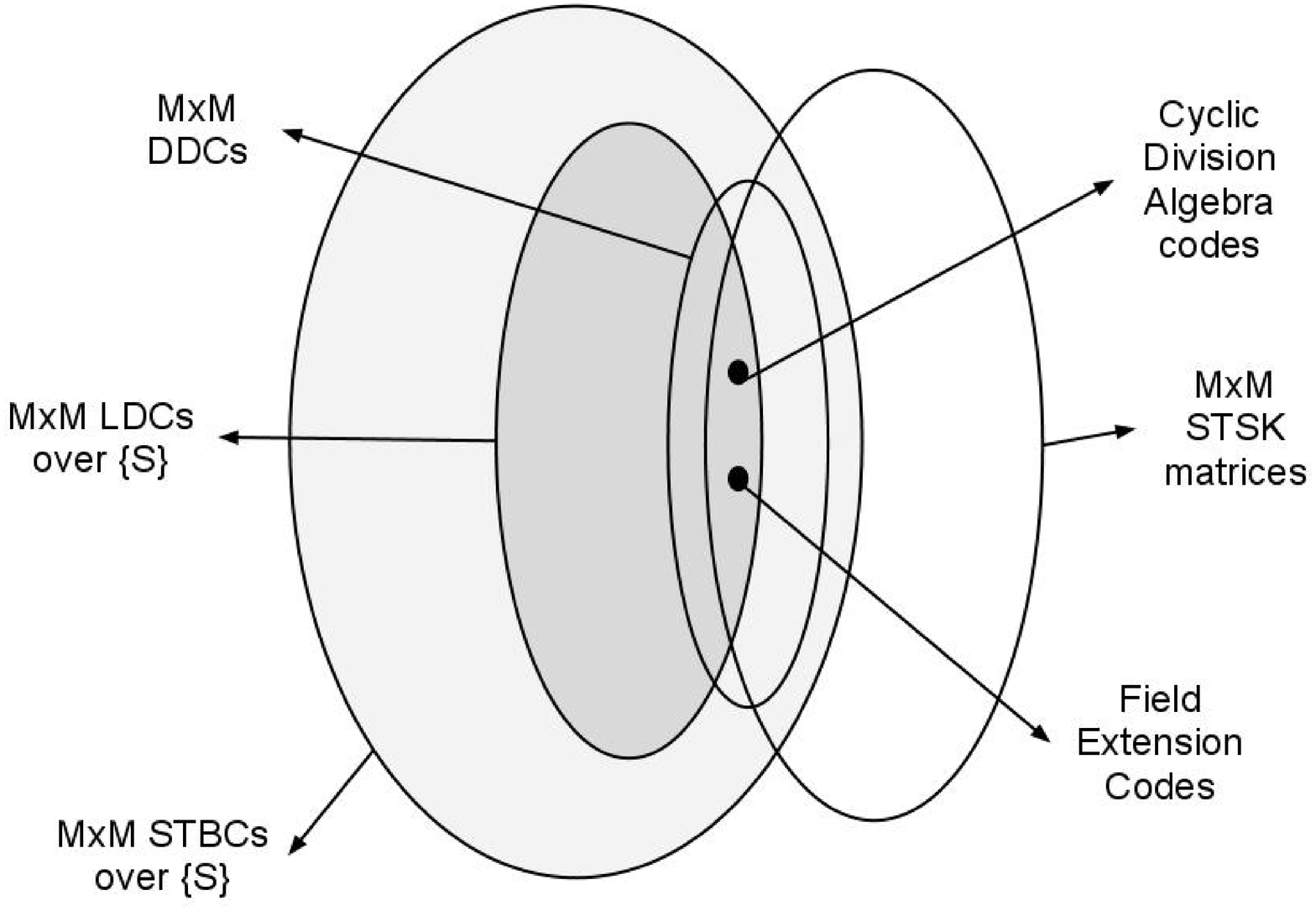}
\caption{Pictorial representation of the connection between the STBCs and the STSK scheme in the space of $M\times M$ matrices over the complex field ${\mathbb C}$.}
\label{VENNDGM}
\end{figure}

Against this background, the following are the novel contributions of this paper:
\begin{enumerate}
 \item We establish a connection between the CSTSK scheme and a class of STBCs termed as Decomposable Dispersion Codes (DDC) and show that the codes from this class result in an ICI-free system. As a result of our established connection, we show that the DDCs enjoy the following benefits that are specific to the STSK scheme:
\begin{itemize}
 \item Low-complexity single-stream based ML detection as described in \cite{LCML}. 
 \item Reduced search-complexity Matched Filtering (MF) based near-ML STSK detection of \cite{SSLCDET}.
\end{itemize}
\item A subclass of Linear Dispersion Codes (LDC) \cite{LDCCODE} is shown in Fig. \ref{VENNDGM} to belong to the class of DDCs. Field Extension Codes (FEC) \cite{BSR1} belong to a subclass of the DDC family, which will be used for deriving structured, full-diversity, high coding gain DMs. However, as the FEC subclass hosts rate-one codes, they are unsuitable for high-rate applications, which motivates us to look for full-rate codes in the broader class of DDCs.
\item Codes from Cyclic Division Algebras (CDA) \cite{BSR2}, \cite{VITERBO} are full-rate, full-diversity, and information lossless codes. We show that these codes also belong to the class of DDCs and hence they enjoy the low-complexity decoding benefits mentioned in 1) above. We then propose a novel method for the systematic construction of DMs from these codes.
\end{enumerate}

Again, Fig. \ref{VENNDGM} depicts the established connection between the STSK scheme, STBCs, and DDCs, and also shows the relationship of the proposed FE and CDA based codes.

The remainder of this paper is organized as follows. In Section II, we briefly describe the STSK signal and our system model, followed by establishing a connection between the STBC and STSK schemes. Section III shows that a subclass of LDCs belongs to the class of DDCs, and a method of obtaining DM set from FE codes is presented using an example. In Section IV, we derive DMs from CDA codes and provide some example constructions considering a PSK signaling set. Section V discusses the code decompositions under non-PSK signal sets, such as square- and star-QAM constellations and their benefits. Section VI discusses the various STSK configurations available for achieving a given rate. Section VII presents our simulation results and discussions, while Section VIII concludes the paper.

{\em Notations:} Boldface uppercase letters represent matrices and are indexed as $\mathbf{X}_i$.  Furthermore, $tr[{\mathbf{X}}]$ and $\mathbf{X}^H$ denote the Trace and Hermitian of the matrix $\mathbf{X}$, respectively. $\mathbf{I}_r$ denotes an $(r\times r)$-element identity matrix. Greek letters like $\zeta$ indicate functions or mappings. Polynomials are represented as a function of $x$, for example $p(x)$. Calligraphic uppercase letters represent sets of matrices, for example $\mathcal{E}$. $\mathcal{D} \subset \mathcal{E}$ implies that $\mathcal{D}$ is a subset of $\mathcal{E}$ and $\lvert\cal{D}\lvert$ represents the cardinality of $\mathcal{D}$. Blackboard-bold font letters like $\mathbb Q$ represent fields. Upper case letters are used to represent sets, fields, and extended fields. The extended field $F=\mathbb{Q}(S)$ represents an
extension of the field of rational numbers $\mathbb Q$ over some set $S$.

\section{CSTSK System and Signal Model}

We consider a MIMO system having $M$ transmit as well as $N$ receive antennas and a quasi-static, frequency-flat fading channel, yielding:

\begin{equation}
 \mathbf{Y}_i=\sqrt{\frac{\mathbf{\rho}}{M}}\mathbf{H}_i\mathbf{X}_i+\mathbf{N}_i,
\label{SYSMOD}
\end{equation}
where $\mathbf{X}_i \in \mathbb C^{M \times T}$ is the transmitted Space-Time (ST) matrix, $\mathbf{Y}_i \in \mathbb C^{N \times T}$ is the received ST matrix, $\mathbf{H}_i \in \mathbb C^{N \times M}$ and $\mathbf{N}_i \in \mathbb C^{N \times T}$ are the channel- and noise-matrices, respectively.  The entries of the channel- and noise-matrices are from a circularly symmetric complex-valued Gaussian distribution i.e., $\cal{CN}$(0,1) and $\cal{CN}$ (0,$N_0$), respectively, where $N_0$ is the noise variance, $\mathbf{\rho}$ is the average Signal to Noise Ratio (SNR) at each receive antenna and $i$ indicates the block index in all the matrices. Throughout this paper we assume $M=T$, that is, we consider only full-diversity, minimum-delay DMs.

For the CSTSK scheme \cite{CSTSK}, we have
\begin{equation}
 \mathbf{X}_i=\mathbf{X}_i^{(q,p)}=s_{i,q}\mathbf{A}_{i,p},
\end{equation}
where $s_{i,q} \in \mathbb C$ is a symbol from an $L$-symbol constellation, $S$, $\mathbf{A}_{i,p} \in \mathbb C^{M \times T}$ is a DM from $\cal{D}$, a set of DMs with $\lvert\cal{D}\lvert$ = $Q$, and $\mathbf{X}_i^{(q,p)} \in \cal{C}$, where $\cal{C}$ is a set of transmitted ST matrices. 
We note that all the DMs $\mathbf{A}_{i,p}$ satisfy the unity average transmission power constraint, i.e.,
\begin{equation}
 tr[{\mathbf{A}_{i,p}}^H{\mathbf{A}_{i,p}}]=T ~\text{for} ~1 \leq p\leq Q.
\label{TEC}
\end{equation}
The notational representation of a typical CSTSK scheme used is formulated as 'CSTSK($M,N,T,Q$), $L$-symbol constellation' \cite{CSTSK}.

\subsection{STSK mapper}
Let $\zeta_p$ be a product-map over a set of ordered pairs, $X=\{ (x_1,x_2) ~|~ x_1 \in X_1, x_2 \in X_2\}$ where $X_1$ and $X_2$ are two arbitrary sets, given by $\zeta_p:(x_1,x_2)\mapsto x_1x_2$. Then, the STSK mapping of a symbol is carried out by applying a DM to the transmitted ST matrix, which is formulated as:
\begin{equation}
 \zeta_p:S\times {\cal D} \mapsto {\cal C}.
\label{PMAP}
\end{equation}
{\em This mapping has to be a one-to-one for the unambiguous detection of the transmitted ST matrices.} Furthermore, it is desirable to have $\operatorname{rank}(\mathbf{X}_i^{(q,p)}-\mathbf{X}_i^{(q^{\prime},p^{\prime})})=M$ for all $p\neq p^{\prime},~\text{or}~q\neq q^{\prime}$ in order to achieve full-diversity, and a high coding gain 
\begin{equation}
 G=\min_{\mathbf{X}_i^{(q,p)}\neq\mathbf{X}_i^{(q^{\prime},p^{\prime})} \in {\cal C}} \Big|\det \Delta \Delta^H\Big|^{1\over M},
\end{equation}
where $~\Delta=(\mathbf{X}_i^{(q,p)}-\mathbf{X}_i^{(q^{\prime},p^{\prime})})$, for the sake of improving the BER performance \cite{TSC}.

\subsection{ICI-free System}
Upon vectorizing Eq.(\ref{SYSMOD}), we arrive at:
\begin{equation}
 \bar{ \mathbf{Y}}_i=\sqrt{\frac{\mathbf{\rho}}{M}}\bar{\mathbf{H}}_i\mathbf{\chi}\mathbf{K}_i+\bar{\mathbf{N}}_i,
\label{SIMSYSMOD}
\end{equation}
where,
\begin{equation}
  \bar{ \mathbf{Y}}_i=vec(\mathbf{Y}_i) \in \mathbb{C}^{NT\times 1},
\end{equation}
\begin{equation}
  \bar{ \mathbf{H}}_i=\mathbf{I}_T \otimes \mathbf{H}_i \in \mathbb{C}^{NT\times MT},
\end{equation}
\begin{equation}
  \bar{ \mathbf{V}}_i=vec(\mathbf{V}_i) \in \mathbb{C}^{NT\times 1},
\end{equation}
\begin{equation}
 \mathbf{\chi}=[vec(\mathbf{A}_1),\hdots, vec(\mathbf{A}_Q)] \in \mathbb{C}^{MT\times Q},
\end{equation}
and
\begin{equation}
 \mathbf{K}_i=[\underbrace{0,\hdots,0}_{p-1},s_{i,q},\underbrace{0,\hdots,0}_{Q-p}] \in \mathbb{C}^{Q\times 1}.
\label{SIMSYSMODEND}
\end{equation}
The equivalent system model of Eq.(\ref{SIMSYSMOD}) is free from ICI, and hence, facilitates both single-antenna based low-complexity ML detection \cite{LCML} and reduced search-complexity MF-based near-ML detection \cite{SSLCDET}.

\subsection{Connection between STBCs and STSK scheme}

{\em Definition 1:} An STBC, $\cal C$, is a finite collection of $(M\times T)$-element matrices with entries from the complex field $\mathbb C$.

{\em Definition 2:} An STBC, $\cal C$, is said be constructed over a signal set $S^{\prime}$, if every element of the codeword matrix is from $\{S^{\prime},S^{\prime *}\}$ or a linear combination of elements of $\{S^{\prime},S^{\prime *}\}$, where $S^{\prime *}$ is a set containing the complex conjugate of the elements of $S^{\prime}$.

{\em Proposition~1:} Any STBC, $\cal{C}$, over a signal set $S^{\prime}$ constitutes an ICI-free system, if there exists a set of matrices $\cal{E}$ such that the map $\zeta_p:S\times {\cal E} \mapsto {\cal C}$ is a bijection, where $S$ is any conventional signal set.
 
{\em Proof:} If there exists a set of $(M\times T)$-element matrices $\cal E$ such that the mapping $\zeta_p:S\times {\cal E} \mapsto {\cal C}$ is a bijection, then we have $\zeta_p^{-1}(\mathbf{X}_k)\neq\zeta_p^{-1}(\mathbf{X}_l)$ for all $\mathbf{X}_k\neq \mathbf{X}_l \in \cal C$, which implies $(s_{i},\mathbf{E}_{j})\neq (s_{i^\prime},\mathbf{E}_{j^\prime})$, where $\mathbf{E}_{j}, \mathbf{E}_{j^\prime} \in \cal E$. This implies that we have either $i\neq i^{\prime} ~\text{or}~ j \neq j^{\prime}$, or both, thus giving us $|{\cal E}|= |{\cal C}|/|S|$. Since we have $Q=|\cal E|$, and $\mathbf{\chi}=[vec(\mathbf{E}_1),\hdots, vec(\mathbf{E}_Q)] $, it is clear from Eq.(\ref{SIMSYSMOD}-\ref{SIMSYSMODEND}) that the STBC is an ICI-free system.

{\em We term this class of STBCs  as Decomposable Dispersion Codes (DDC).}
In the following section we will show that as seen in Fig. {\ref{VENNDGM}} a subclass of LDCs belongs to the class of DDCs and present examples of both rate-one and full-rate codes of this class.

\section{Proposed LDC based Decomposable Dispersion Codes}
In this section we first show that a special class of LDCs belongs to the family of DDCs. Then, a systematic method conceived for obtaining DMs from this special class of LDCs is presented. Furthermore, FECs are shown to be a member of such an LDC class, and an example DM set construction based on FECs is constructed for CSTSK(2,2,2,4), 4-PSK system.

An LDC is defined by a set of matrices of the form 
\begin{equation}
  \mathcal{C}=\left\lbrace \sum_{i=0}^{V-1}f_i\mathbf{M}_i+f^{*}_i\mathbf{M^{\prime}}_i ~\Big|~ f_i \in S, i=0,1,\cdots,M-1 \right\rbrace,
\end{equation}
where $S$ is an arbitrary signal set, $V$ is the number of substreams of the data sequence, while $\mathbf{M}_i$ and $\mathbf{M^\prime}_i$ are $(M \times T)$-element DMs, which were chosen by maximizing the mutual information in \cite{LDCCODE}. 

We consider a special class of LDCs in which 
\begin{itemize}
 \item the DMs $\mathbf{M^\prime}_i$ are zero matrices, and
 \item $S$ is an arbitrary PSK signal set,
\end{itemize}
which gives
\begin{equation}
 {\cal C}={\left\lbrace \sum_{i=0}^{V-1}f_i {\mathbf M}_i ~\Big|~\forall~ f_i\in S \right\rbrace}.
\label{LDCEQN}
\end{equation}
Different codebooks of this form are characterized by different sets of DMs $\{\mathbf{M}_i\}_{i=0}^{V-1}$. For example, $\{\mathbf{M}_i\}_{i=0}^{V-1}$ can be chosen by maximizing the coding gain or mutual information over large sets of complex-valued Gaussian random matrices. They can be powers of matrices constructed by Field Extensions \cite{BSR1}, or they can be chosen based on frame-theoretic considerations which result in maximum capacity- or maximum coding gain DMs \cite{HEATH}. 

{\em Theorem~1:} If $\cal{C}$ is an LDC as defined in Eq.(\ref{LDCEQN}) then the mapping becomes
\begin{equation}
 \zeta_p:S \times {\cal E}\mapsto {\cal C}
\label{DDCMAP}
\end{equation}
where 
\begin{equation}
{\cal E}={\left\lbrace \mathbf{M}_l+\sum_{i=0,i\neq l}^{V-1}f_i^\prime \mathbf{M}_i ~\Big|~ \forall f_i^\prime \in S \right\rbrace}~\text{for any}~ 0 \leq l \leq V-1, \label{ELEQDDC}
\end{equation}
is a bijection.

{\em Proof:} Please refer to Appendix B. 

{\em Corollary~1:} From {\em Theorem~1} and {\em Proposition~1}, it is straightforward that to show for $S=S^{\prime}$ the LDCs given in Eq.(\ref{LDCEQN}) are DDCs. Thus, they enjoy the low-complexity detection benefits of the STSK schemes of \cite{LCML}, \cite{SSLCDET}.

We can now infer from Eq.(\ref{DDCMAP}) and Eq.(\ref{ELEQDDC}) that $|{\cal C}|=|S|\cdot |{\cal E}|=L\cdot L^{V-1}$ and any ${\cal D} \subseteq {\cal E}$
can be used as a set of DMs. Thus, the number of DMs that may be generated from LDC based DDCs is $1 \leq Q \leq L^{V-1}$.

\subsection{Systematic selection of subsets of $\mathcal{E}$}
Let ${\cal L}_r=\{0,1,2,\hdots,r-1\}\subset \{i\}_{i=0}^{V-1}$ for $1\leq r \leq V-1$. If we have \[{\mathcal{E}}_{{\cal{L}}_r}={\left\lbrace \sum_{i \in {{\cal{L}}_r}}\mathbf{M}_i+\sum_{i \in \bar{{\cal{L}}_r}}f_i^\prime{\mathbf{M}}_i ~|~ f_i^\prime \in S \right\rbrace},\] where $\bar{\cal L}_r$ is the complement of ${\cal L}_r$, then it is straightforward to show that ${\mathcal{E}}_{{\cal{L}}_{V-1}}\subset \cdots \subset {\mathcal{E}}_{{\cal{L}}_2} \subset {\mathcal{E}}_{{\cal{L}}_1} \subset {\cal C}_{DDC}$ and $|{\mathcal{E}}_{{\cal{L}}_r}|=L^{V-r}$. Thus, when we have ${\cal D}={\mathcal{E}}_{{\cal{L}}_{r}}$, then the rate achieved by the STSK scheme is
\begin{equation}
 R_{STSK-DDC}=\frac{(V-r+1)\log_2{L}}{T}~\text{bpcu}.
\label{RLDC}
\end{equation}
Hence, for a desired rate $R$ with a fixed $L$ we
\begin{itemize}
 \item obtain $r$ from Eq.(\ref{RLDC}), 
 \item find the corresponding set ${\cal L}_r$, and
 \item get the DM set ${\cal D}={\cal E}_{{\cal L}_r}$.
\end{itemize}

\subsection{DM set construction based on Field Extension Codes}
Let $S$ denote the signal set over which FECs are constructed. We restrict $S$ to be an arbitrary PSK constellation having $|S|=L$.

Let $F$ be a number field, given by $\mathbb{Q}(S)$, and $K$ be an $M^{th}$ degree algebraic extension of $F$ over
$\alpha$, which is formulated as, $K=F(\alpha)$ such that $p(\alpha)=0$, where the irreducible monic polynomial $p(x) \in F[x]
~\text{is given by}~ p(x)=x^n+a_{n-1}x^{n-1}+\cdot\cdot\cdot+a_1x+a_0$. Thus, we have the following chain of field extensions.
\begin{equation}
\mathbb{Q} \subset \mathbb{Q} (S)=F \subset \mathbb{Q} (S,\alpha) = F(\alpha)=K,
\label{FE1}
\end{equation}
where any $k\in K$ can be represented as $k= \sum_{i=0}^{M-1}f_i{\alpha}^i$, where $f_i$ are from $F$. It was shown in \cite{BSR1} that $\exists$ a natural mapping for all $k\in K$, given by $k\mapsto \lambda_k$, where $\lambda_k$ maps any $u\in K$ to $ku$. The unique matrix associated with $\lambda_k$ is
given by ${ \sum_{i=0}^{M-1}f_i {\mathbf M}^i}$ \cite{BSR1}, \cite{BSR2} where,
\begin{equation}
\mathbf{M}=\left[\begin{array}{ccccc}
 0 &  0 &  \cdots & 0 & -a_0\\
 1 &  0 &  \cdots & 0 & -a_1\\
 0 &  1 &  \cdots & 0 & -a_2\\
 \vdots & \vdots & \ddots & \vdots & \vdots\\
0 &  0 & \cdots & 1 & -a_{n-1}\\\end{array}
\right].
\label{M1}
\end{equation}
Thus, the resultant FEC is given by
\begin{equation}
{\cal C}={\left\lbrace \sum_{i=0}^{M-1}f_i {\mathbf M}^i ~\Big|~\forall~ f_i\in F \right\rbrace}.
\label{FECCODE}
\end{equation}

Thus, by assuming $F=S$ in Eq.(\ref{FECCODE}) and exploiting {\em Theorem~1}, we can write $\zeta_p: S\times {\cal E} \mapsto {\cal C}$, where $\zeta_p$ is a bijection.
Any, ${\cal D} \subseteq {\cal E}$ can be used as a set of DMs. Furthermore, we have $|{\cal C}|=L^M=|S|\cdot|{\cal E}|=L\cdot L^{M-1}$,
thus, $1\leq Q\leq L^{M-1}$.

{\em Example~1:} When considering $L=4$, $M=2$, $x^2-e^{j\frac{2\pi}{4}}$ becomes irreducible over $F$ \cite{BSR2}. Thus for $F=S=\{1,-1,j,-j\}$ and $l=1$,
from eqn.(\ref{ELEQDDC}) we have
\[
 \mathcal{E}= \left\lbrace\left[\begin{array}{cc} 1 & j\\ 1 & 1\\\end{array}\right]
	      \left[\begin{array}{cc} j & j\\ 1 & j\\\end{array}\right]
	      \left[\begin{array}{cc} -1 & j\\ 1 & -1\\\end{array}\right]
		\left[\begin{array}{cc} -j & j\\ 1 & -j\\\end{array}\right] \right\rbrace.
\]

We can choose $\mathcal{D}$ to be any subset of $\mathcal{E}$ upon normalizing it by $\sqrt{1\over M}$, which is $\sqrt{1\over2}$ in the above case to satisfy Eq.(\ref{TEC}). Given $\mathcal{D}=\mathcal{E}$ we get four DMs. The coding gain of this scheme may be shown to be $G=1$. We refer to these DMs as Field Extension Code based Dispersion Matrices (FEC-DM). 

\section{Proposed full-rate CDA code based Decomposable Dispersion Code}

In this section we show that the codes from CDAs are DDCs and hence they may be used for STSK schemes. We propose a method for obtaining DMs from CDA codes for achieving a desired rate and present a construction example for the CSTSK(2,2,2,8), BPSK system.

We consider CDA codes from transcendental extensions of $\mathbb{Q}$ \cite{CDA1}, as they do not depend on the number of antennas or on the signaling set, while offering a better coding gain than the CDA codes constructed from cyclotomic extensions of $\mathbb{Q}$.

Considering codes constructed from CDAs over the field $F=\mathbb{Q}(S,t,\omega_M)$, we get the full-diversity, full-rate $(M\times M)$-element Space-Time (ST) codes \cite{VITERBO}, \cite{CDA1} given by 
\begin{equation}
\hspace{-10pt}
\footnotesize
{\cal{C}}=\left\{\left[\begin{array}{ccccc}
 \sum_{i=0}^{M-1}f_{0,i}(t_M)^i & \delta \sigma(\sum_{i=0}^{M-1}f_{M-1,i}(t_M)^i)  & \cdots & \delta \sigma^{M-1}(\sum_{i=0}^{M-1}f_{1,i}(t_M)^i)\\
  \sum_{i=0}^{M-1}f_{1,i}(t_M)^i &  \sigma(\sum_{i=0}^{M-1}f_{0,i}(t_M)^i) &  \cdots & \delta \sigma^{M-1}(\sum_{i=0}^{M-1}f_{2,i}(t_M)^i)\\
 \sum_{i=0}^{M-1}f_{2,i}(t_M)^i &  \sigma(\sum_{i=0}^{M-1}f_{1,i}(t_M)^i)  & \cdots & \delta \sigma^{M-1}(\sum_{i=0}^{M-1}f_{3,i}(t_M)^i)\\
 \vdots & \vdots & \ddots & \vdots\\
\sum_{i=0}^{M-1}f_{M-1,i}(t_M)^i & \sigma(\sum_{i=0}^{M-1}f_{M-2,i}(t_M)^i) & \cdots &  \sigma^{M-1}(\sum_{i=0}^{M-1}f_{0,i}(t_M)^i)\\\end{array}
\right] \Big| f_{i,j} \in S ~ \text{for}~  0 \leq i,j \leq M-1 \right\},
\label{STMat1}
\end{equation}
\begin{equation}
\footnotesize
=\left\{\left[\begin{array}{ccccc}
 \sum_{i=0}^{M-1}f_{0,i}(t_M)^i &  \delta\sum_{i=0}^{M-1}f_{M-1,i}(\omega_Mt_M)^i  & \cdots &  \delta\sum_{i=0}^{M-1}f_{1,i}(\omega_M^{M-1}t_M)^i\\
  \sum_{i=0}^{M-1}f_{1,i}(t_M)^i &  \sum_{i=0}^{M-1}f_{0,i}(\omega_Mt_M)^i &  \cdots &  \delta\sum_{i=0}^{M-1}f_{2,i}(\omega_M^{M-1}t_M)^i\\
 \sum_{i=0}^{M-1}f_{2,i}(t_M)^i &  \sum_{i=0}^{M-1}f_{1,i}(\omega_Mt_M)^i  & \cdots &  \delta\sum_{i=0}^{M-1}f_{3,i}(\omega_M^{M-1}t_M)^i\\
 \vdots & \vdots & \ddots & \vdots\\
\sum_{i=0}^{M-1}f_{M-1,i}(t_M)^i & \sum_{i=0}^{M-1}f_{M-2,i}(\omega_Mt_M)^i & \cdots &  \sum_{i=0}^{M-1}f_{0,i}(\omega_M^{M-1}t_M)^i\\\end{array}
\right] \Big| f_{i,j} \in S ~ \text{for}~  0 \leq i,j \leq M-1 \right\},
\label{STMat2}
\end{equation}
where $\sigma$ is the Galois group generator that fixes $f_{i,j}$ and maps $t_M$ to $\omega_Mt_M$, while the transcendental elements $t$ and $\delta$ are chosen from the unit circle to achieve information losslessness\footnote{An STBC is said to be {\em information lossless} if its generator matrix is unitary. For more details refer to \cite{INFOLL}.} \cite{CDA1}. Furthermore, we have $|{\cal C}|=L^{M^2}$ as $\cal C$ is a full-rate code. 

For the ease of presentation, we adopt the following notation for describing the set in Eq.(\ref{STMat2}):
\begin{equation}
\cal C= \left[\begin{array}{ccccc}
        {\hat{K}}^{(0,0)} & {\delta\hat{K}}^{(M-1,1)} & \cdots & {\delta\hat{K}}^{(1,M-1)}\\
	{\hat{K}}^{(1,0)} & {\hat{K}}^{(0,1)} & \cdots & {\delta\hat{K}}^{(2,M-1)}\\
	\vdots & \vdots & \ddots & \vdots\\
	{\hat{K}}^{(M-1,0)} & {\hat{K}}^{(M-2,1)} & \cdots & {\hat{K}}^{(0,M-1)}\\
       \end{array}
        \right]\label{SIMPC1},
\end{equation}
where, we have
\begin{equation*}
 \hat K^{(j,k)}=\left\{ \sum_{i=0}^{M-1}f_{j,i}({\omega_M}^kt_M)^i \Big| f_{j,i} \in S ~ \text{for}~  0 \leq i,j \leq M-1 \right\}.
\end{equation*}
 
In $\hat K^{(j,k)}$ of Eq.(\ref{SIMPC1}), the superscript $j$ captures the $M$ distinct sets containing $M$ independent
symbols each, i.e., $\{f_{j,i}\}_{i=0}^{M-1}$, and the superscript $k$ is the distinct index of the coefficients of
$\{f_{j,i}\}_{i=0}^{M-1}$ associated with each column in Eq.(\ref{STMat2}).

{\em Proposition~2:} A CDA code constructed over an arbitrary PSK signal results in an ICI-free system, which hence may be viewed as a STSK scheme. Thus, the CDA codes enjoy the low-complexity detection benefits of the STSK scheme \cite{LCML}, \cite{SSLCDET}.

\begin{proof}
We present the proof in two steps. In Step I we consider the diagonal elements of the CDA code. In Step II, under a bijective product mapping we achieve the decomposition of the off-diagonal elements and hence the complete CDA code. We conclude the proof by invoking {\em Proposition~1}.

Step I:
Let $F$ be an algebraic number field defined by $\mathbb{Q}(S,t,\omega_M)$, where $t$ is a transcendental element over $\mathbb{Q}(S)$, and $\omega_M=e^{j\frac{2\pi}{M}}$. Let $K$ be an $M^{th}$ degree algebraic extension of $F$ over $t_M=t^{{1}/{M}}$, i.e. $K=F(t^{{1}/{M}})$.
Thus, we can write 
\begin{equation}
K=\left\lbrace\sum_{i=0}^{M-1}f_i{(t_M)^i} ~\Big|~ f_i\in F ~\text{for}~ i=0,1,\cdots,M-1 \right\rbrace.
\label{KEQ}
\end{equation}

{\em Theorem~2:} Let $S$, $F$ and $K$ be defined as above. Let $\zeta_p$ be a product mapping as defined earlier.
If $F=S$ in Eq.(\ref{KEQ}), then the map $\zeta_p: S\times K_l \mapsto K$, where we have
\begin{equation}
\hspace{-10 pt}
K_l={\left\lbrace (t_M)^l+\sum_{i=0,i\neq l}^{M-1}f_i^\prime({t_M})^i ~\Big|~ f_i^\prime \in S,~ i=0,1,\hdots,M-1 \right\rbrace}
\label{KLEQ}
\end{equation}
for any $0\leq l \leq M-1$, is a bijection.

{\em Proof:} The proof of {\em Theorem~1} given in Appendix B holds, when  $\mathbf{M}_i$, $V$, ${\cal E}$, ${\cal E}^{\prime}$, and ${\cal C}$ are replaced by $(t_M)^i$, $M$, $K_l$, $K_l^{\prime}$, and $K$, respectively.

Applying {\em Theorem~2} to the set along the main diagonal of $\cal C$ in Eq.(\ref{SIMPC1}), we arrive at
\begin{equation}
\small
\hspace{-20pt}
{
\cal C= \left[\begin{array}{ccccc}
        {\zeta_p(S\times \hat{K}}^{(0,0)}_{l_1}) & {\delta\hat{K}}^{(M-1,1)} & \cdots & {\delta\hat{K}}^{(1,M-1)}\\
	{\hat{K}}^{(1,0)} & {\zeta_p(S\times \hat{K}}^{(0,1)}_{l_2}) & \cdots & {\delta\hat{K}}^{(2,M-1)}\\
	\vdots & \vdots & \ddots & \vdots\\
	{\hat{K}}^{(M-1,0)} & {\hat{K}}^{(M-2,1)} & \cdots & {\zeta_p(S\times \hat{K}}^{(0,M-1)}_{l_M})\\
       \end{array}
        \right],\label{SIMPC2}}
\end{equation}
where, we have
\begin{equation*}
\hspace{-35 pt}
 \hat K^{(j,k)}_{l_m}=\Big\{({\omega_M}^kt_M)^{l_m}+ \sum_{i=0,i\neq l_m}^{M-1}f_{j,i}({\omega_M}^kt_M)^i
\end{equation*}
\begin{equation*}
\hspace{15 pt}
\hspace{100pt} \Big| f_{j,i} \in S ~ \text{for}~  0 \leq i,j \leq M-1 \Big\}
\end{equation*}
for $0\leq \{l_m\}_{m=1}^{M} \leq M-1$.

Step II: With the aid of {\em Theorem~3} we will show below that the off-diagonal sets in Eq.(\ref{SIMPC1}) can be decomposed into the product of two sets.

{\em Theorem~3:} Let $S$, $K$, $\zeta_p$, and $F$ be defined as before. If we have $F=S$ in Eq.(\ref{KEQ}), then
$\zeta_p$:$~S\times K \mapsto K$ is a bijection.

{\em Proof:} Please refer to Appendix C.

Applying {\em Theorem~3} to the off-diagonal elements of Eq.(\ref{SIMPC2}), we arrive at Eq.(\ref{SIMPC3}).
\begin{equation}
{
\cal C= \left[\begin{array}{ccccc}
        {\zeta_p(S\times \hat{K}}^{(0,0)}_{l_1}) & {\zeta_p(S\times \delta\hat{K}}^{(M-1,1)}) & \cdots & {\zeta_p(S\times \delta\hat{K}}^{(1,M-1)})\\
	{\zeta_p(S\times \hat{K}}^{(1,0)}) & {\zeta_p(S\times \hat{K}}^{(0,1)}_{l_2}) & \cdots & {\zeta_p(S\times \delta\hat{K}}^{(2,M-1)})\\
	\vdots & \vdots & \ddots & \vdots\\
	{\zeta_p(S\times \hat{K}}^{(M-1,0}) & {\zeta_p(S\times\hat{K}}^{(M-2,1)}) & \cdots & {\zeta_p(S\times \hat{K}}^{(0,M-1)}_{l_M})\\
       \end{array}
        \right]\label{SIMPC3}}
\end{equation}
Thus, from Eq.(\ref{SIMPC3}) we generate the bijective mapping $ \zeta_p$:$~S\times{\cal E} \mapsto {\cal C}$, where,
\begin{equation}
{
\cal E= \left[\begin{array}{ccccc}
        { \hat{K}}^{(0,0)}_{l_1} & {\delta\hat{K}}^{(M-1,1)} & \cdots & {\delta\hat{K}}^{(1,M-1)}\\
	{\hat{K}}^{(1,0)} & { \hat{K}}^{(0,1)}_{l_2} & \cdots & {\delta\hat{K}}^{(2,M-1)}\\
	\vdots & \vdots & \ddots & \vdots\\
	{\hat{K}}^{(M-1,0)} & {\hat{K}}^{(M-2,1)} & \cdots & { \hat{K}}^{(0,M-1)}_{l_M}\\
       \end{array}
        \right],\label{SIMPC4}}
\end{equation}
and hence, from {\em Proposition~1} we conclude that the CDA codes result in an ICI-free system.
\end{proof}

From Eq.(\ref{SIMPC3}), we can infer that $|{\cal C}|=L^{M^2}=|S|\cdot|{\cal E}|=L\cdot L^{M^2-1}$ and any set ${\cal D}\subseteq {\cal E}$ can be used as a set of DMs. Thus, CDA codes offer $Q$ number of DMs, where we have $1\leq Q\leq L^{M^2-1}$. By contrast, for FEC-DMs we have $1\leq Q\leq L^{M-1}$. We refer to these DMs obtained from the CDA codes as Cyclic Division Algebra code based DMs (CDA-DM).

{\em Example~2:} Let $S=\lbrace1,-1\rbrace$, and $t$ as well as $\delta$ be chosen from within the unit circle. Let furthermore the number of transmit antennas be $M=2$, and $l_1=l_2=1$.
From Eq.(\ref{SIMPC4}), we get
\begin{equation}
{
\mathcal{E}=\left\lbrace \left[\begin{array}{cc}
 1+f_{0,1}t_2 &  \delta(f_{1,0}-f_{1,1}t_2)\\
 f_{1,0}+f_{1,1}t_2 &  (1-f_{0,1}t_2)\\\end{array}
\right] \Big| ~f_{i,j} \in S \right\rbrace
}
\end{equation}
with $\lvert{\cal E}\lvert = \lvert{S}\lvert$ $^3$ = 8. In order to satisfy the unit average transmission energy constraint of Eq.(\ref{TEC}), the matrices in the set $\cal E$ are scaled by $1\over 2$ (in general $1 \over M$) to arrive at
\begin{equation}
{
\mathcal{D}=\left\lbrace {1\over 2}\left[\begin{array}{cc}
 1+f_{0,1}t_2 &  \delta(f_{1,0}-f_{1,1}t_2)\\
 f_{1,0}+f_{1,1}t_2 &  (1-f_{0,1}t_2)\\\end{array}
\right] \Big| ~f_{i,j} \in S \right\rbrace
}
\end{equation}
with $Q=8$.

\subsection{Systematic selection of subsets of $\cal{E}$}
Let $K$, $S$, and $K_l$ be defined as before. Let ${\cal L}_r=\{0,1,2,\hdots,r-1\}\subset \{i\}_{i=0}^{M-1}$ 
for $1\leq r \leq M-1$.
If $\hat{K}_{{\cal{L}}_r}^{(j,k)}={\left\lbrace \sum_{i \in {{\cal{L}}_r}}({\omega_M}^kt_M)^i+\sum_{i \in \bar{{\cal{L}}_r}}f_{j,i}{({\omega_M}^kt_M)}^i ~\big|~ f_{(j,i)} \in S \right\rbrace}$,
where $\bar{{\cal{L}}_r}$ is the complement of ${{\cal{L}}_r}$, then it is straightforward to show that
\begin{equation*}
{\hat{K}}^{(j,k)}_{{\cal{L}}_{M-1}}\subset {\hat{K}}^{(j,k)}_{{\cal{L}}_{M-2}} \subset \cdots \subset {\hat{K}}^{(j,k)}_{{\cal{L}}_2} \subset {\hat{K}}^{(j,k)}_{{\cal{L}}_1} = {\hat{K}}^{(j,k)}_{1}\subset {\hat{K}}^{(j,k)},
\end{equation*}
and $\lvert{\hat{K}}^{(j,k)}_{{\cal{L}}_r} \lvert = L^{M-r}~\forall ~0\leq j,k \leq M-1$. Let $\hat{K}_{\cal L}^{(j,k)}=\{{\hat{K}}^{(j,k)}_{{\cal{L}}_1},{\hat{K}}^{(j,k)}_{{\cal{L}}_2},\hdots,{\hat{K}}^{(j,k)}_{{\cal{L}}_{M-1}}\}$.
Then, from {\em Theorem~2} and {\em Theorem~3}, we arrive at the one-to-one mapping $\zeta_p:S\times{\cal E}_r \mapsto {\cal C}_r \subseteq {\cal C}$ where, we have
\begin{equation}
{
{\cal E}_r= \left[\begin{array}{ccccc}
        { \hat{K}}^{(0,0)}_{{\cal L}_r} & {\delta\hat{K}}^{(M-1,1)} & \cdots & {\delta\hat{K}}^{(1,M-1)}\\
	{\hat{K}}^{(1,0)} & { \hat{K}}^{(0,1)}_{{\cal L}_r} & \cdots & {\delta\hat{K}}^{(2,M-1)}\\
	\vdots & \vdots & \ddots & \vdots\\
	{\hat{K}}^{(M-1,0)} & {\hat{K}}^{(M-2,1)} & \cdots & { \hat{K}}^{(0,M-1)}_{{\cal L}_r}\\
       \end{array}
        \right].}
\end{equation}
Thus, we have $|{\cal C}_r|=L^{M^2-r+1}=|S|\cdot|{\cal E}_r|=L\cdot L^{M^2-r}$ and
for $r=1$, we get ${\cal C}_r={\cal C}$. Further generalizing this, we get
\begin{equation}
 \zeta_p:S\times{\cal E}_{(m,r)} \mapsto {\cal C}_{(m,r)} \subseteq {\cal C},
\label{GEN2MAP}
\end{equation}
where,
\begin{equation}
{
{\cal E}_{(m,r)}= \left[\begin{array}{ccccc}
        { \hat{K}}^{(0,0)}_{{\cal L}_r} & {\delta\hat{K}}^{(M-1,1)} & \cdots & {\delta\hat{K}}^{(1,M-1)}_{{\cal L}_r}\\
	{\hat{K}}^{(1,0)}_{{\cal L}_r} & { \hat{K}}^{(0,1)}_{{\cal L}_r} & \cdots & {\delta\hat{K}}^{(2,M-1)}_{{\cal L}_r}\\
	\vdots & \vdots & \vdots & \vdots\\
	{\hat{K}}^{(m-1,0)}_{{\cal L}_r} & { \hat{K}}^{(m-2,1)}_{{\cal L}_r} & \cdots & {\delta\hat{K}}^{(m,M-1)}\\
	{\hat{K}}^{(m,0)} & { \hat{K}}^{(m-1,1)}_{{\cal L}_r} & \cdots & {\delta\hat{K}}^{(m+1,M-1)}\\
	\vdots & \vdots & \vdots & \vdots\\
	{\hat{K}}^{(M-1,0)} & {\hat{K}}^{(M-2,1)} & \cdots & { \hat{K}}^{(0,M-1)}_{{\cal L}_r}\\
       \end{array}
        \right],}
\end{equation}
for $1\leq m\leq M$, and ${\cal C}_{(m,r)}$ is the image of $\zeta_p$ in ${\cal C}$.
Here, $|{\cal C}_{(m,r)}|=L^{M^2-mr+1}=|S|\cdot|{\cal E}_{(m,r)}|=L\cdot L^{M^2-mr}$ and
for $r=1$ and $m=1$, we get ${\cal C}_{(m,r)}={\cal C}$.

It becomes clear from the above analysis that ${\cal E}_{(m,r)}$ relying on any legitimate $(m,r)$ can be chosen as $\cal D$ that gives $Q=L^{M^2-mr}$, thus achieving an effective throughput of 
\begin{equation}
R_{STSK-CDA}=\frac{(M^2-mr+1)\log_2{L}}{T}~\text{bpcu}.
\label{RCDA}
\end{equation}

Hence, for a desired rate $R$ and fixed $L$ we
\begin{itemize}
 \item obtain a legitimate pair $(m,r)$ from Eq.(\ref{RCDA}), 
 \item find the corresponding set ${\cal L}_r$, and
 \item get the DM set ${\cal D}={\cal E}_{(m,r)}$.
\end{itemize}

\section{Decomposition of FECs and CDA codes over QAM signal sets}

The proposed FECs and CDA codes were shown to be decomposable over arbitrary PSK signal sets in Sections III and IV. In this section we show that these codes are decomposable over QAM constellations as well, and discuss the suitability of star-QAM \cite{STARQAM} versus square-QAM signal sets for reduced search complexity, when decoding these STBCs.

Consider the subclass of LDCs $\cal C$ of Eq.(\ref{LDCEQN}) considered with a QAM signal set $S$ instead of a PSK constellation, where we have, ${\cal C}={\left\lbrace \sum_{i=0}^{V-1}f_i {\mathbf M}_i ~\Big|~\forall~ f_i\in S \right\rbrace}$ with $S$ being a square QAM signal set. Let $|S|=L=4^a$, where $a$ is any positive integer, which ensures that the QAM constellations are of square type. Let $S_{L/4}=\{s_1,\hdots,s_{L/4}\}$ be the ordered set of $L$-QAM points belonging to the first quadrant. Hence, for all $s_i\in S_{L/4}$, we have $\Re \{ s_i \} ~\text{and}~ \Im \{ s_i\} > 0$. Let $g=e^{j\frac{\pi}{2}}$, $S^{\prime}=\{s_i g^{m_i},~1 \leq i \leq L/4 |~m_i \in [0, 3]\}\subset S$ and $S_{sym}=\{1,g,g^2,g^3\}$. Thus it is easy to see that we have $S=\{S^{\prime}, S^{\prime}g, S^{\prime}g^2, S^{\prime}g^3 \}$, and hence any element in $S$ can be uniquely written as $s_ig^{k}$, where $s_i\in S_{L/4}$ and $0\leq k\leq 3$.

{\em Theorem~4:} Let $S$, $S^{\prime}$, $S_{sym}$, and $\cal C$ be defined as above. Then the mapping
\begin{equation}
 \zeta_p:S_{sym} \times {\cal E}\mapsto {\cal C}
\label{}
\end{equation}
where 
\begin{equation}
{\cal E}={\left\lbrace f_l\mathbf{M}_l+\sum_{i=0,i\neq l}^{V-1}f_i^\prime \mathbf{M}_i ~\Big|~ f_l \in S^{\prime},~\forall f_i^\prime \in S \right\rbrace}~ \label{LDCQAMDEC}
\end{equation}
for any $0 \leq l \leq V-1$ is a bijection.

{\em Proof:} Please refer to Appendix D.

{\em Corollary~2:} From {\em Proposition~1} and {\em Theorem~4} it is straightforward to show that the LDCs given in Eq.(\ref{LDCEQN}) constructed over square-QAM signal sets are decomposable over $\{1,-1,j,-j\}$ and hence result in an ICI-free system.

It is evident from Eq.(\ref{LDCQAMDEC}) that $|{\cal E}|=(L/4) \cdot L^{V-1}=|{\cal C}|/4$. Thus, with the aid of the detector of \cite{SSLCDET} the search complexity of these codes can be reduced from $|{\cal C}|$ to $|{\cal C}|/4$. 

Consider now the subclass of LDCs $\cal C$ of Eq.(\ref{LDCEQN}) in conjunction with star-QAM signal set \cite{STARQAM} instead of a PSK constellation, i.e, ${\cal C}={\left\lbrace \sum_{i=0}^{V-1}f_i {\mathbf M}_i ~\Big|~\forall~ f_i\in S \right\rbrace}$, where $S$ represents the $L$-star-QAM constellation with $L=|S|=4^a$ and $2^{a-1}$ rings in the constellation. Let $S_{amp}=\{s_1,\hdots,s_{2^{a-1}}\}$ represent the ordered set of $2^{a-1}$ real and positive signal points on the real axis of the constellation. Let furthermore $g=e^{j\frac{\pi}{2^a}}$, $S_{sym}=\{g^q|~0\leq q\leq 2^{a+1}-1\}$, and $S^{\prime}=\{s_ig^{m_i},1\leq i\leq 2^{a-1} |~m_i\in [0,2^{a+1}-1]\}$. It can be verified that {\em Theorem~4} holds for the sets $S$, $S^{\prime}$, $S_{sym}$, $S_{L/4}=S_{amp}$, and ${\cal C}$ as defined above. Thus, the decoding complexity of the star-QAM constellation \cite{STARQAM} is given by $\max \{|S_{amp}|,|{\cal E}| \}=\max{\{2^{a+1},\frac{|{\cal C}|}{2^{a+1}}\}}=\frac{|{\cal C}|}{2^{a+1}}$ for $V \geq 2$ and $a \geq 1$. Since $a \geq 1$, $U=2^{a+1}\geq 4$. Thus, using a star-QAM constellation instead of square-QAM constellation has the following benefits:
\begin{itemize}
 \item Star-QAM offers a higher number of axes of symmetry in the constellation that enables further reduction in search complexity from $|{\cal C}|/4$ to $|{\cal C}|/U$, where $U\geq4$ as given above.
 \item The performance of the STSK scheme combined with star-QAM is better than that with the square-QAM constellation under both ML and low complexity detectors \cite{SSLCDET}.
\end{itemize}

For example, FECs designed for two transmit antenna aided 16-square-QAM gives $\cal C$ and $\cal E$ such that $|{\cal C}|=256$ and ${|\cal E|}=64$, since, $|S_{sym}|=4$. Thus, the search complexity of the detector of \cite{SSLCDET} is 64. Considering 16-star-QAM instead of 16-square-QAM we get $|{\cal E}|=32$, since $|S_{sym}|=8$, which reduces the complexity from 64 to 32. Thus, it is desirable to use constellations with a higher number of axes of symmetry in order to reduce the decoding complexity.

By applying {\em Theorem~4} to the diagonal elements of Eq.(\ref{SIMPC1}) and following the lines of the proof of {\em Proposition~2} it can be shown that the CDA codes also decompose over square-QAM and star-QAM constellations. Thus, with the aid of the low complexity detector of \cite{SSLCDET} both the CDA codes as well as the so-called Perfect Space-Time Codes \cite{VITERBO} can be decoded at a reduced complexity. However, as a benefit of our established connection between STSK and STBCs, all the DDCs - including Perfect Space-Time Codes - may use the SM-specific low-complexity sphere decoding (SD) techniques of \cite{MDRENZO1}, \cite{MDRENZO2}. However, in the rest of the paper we restrict our discussions to FECs and CDA codes designed for PSK signal sets, for which SD is not applicable, and leave the code constructions for square- and star-QAM signal sets as well as their decoding for future research.

\section{Choice of Dispersion Matrices for different STSK configurations}

In this section we discuss the various design configurations available for achieving a desired rate in an STSK scheme with CO-, FEC-, and CDA-DMs. For example, consider CSTSK(2,2,2,$Q$) system, with the desired transmission rate of $R$ bpcu. Table II illustrates the various possibilities of $Q$ and $L$ for $R=1,1.5,2,2.5$ bpcu. For each of the options available for a given rate in Table II, the corresponding $Q$ number of CO-DMs are obtained by maximizing the mutual information over a large set of complex-valued matrices with Gaussian random entries. For a given $Q_{desired}$, the number of DMs based on FECs and CDA codes have an upper limit, which depends on the constellation size $L$, since we have $1\leq Q\leq L^{M-1}$ for FEC-DMs, and $1\leq Q \leq L^{M^2-1}$ for CDA-DMs. One way to overcome this limitation is to consider the smallest possible $L^{\prime}$ such that $L^{\prime M-1} \geq Q_{desired}$ in case of FEC-DMs and $L^{\prime M^2-1}\geq Q_{desired}$ in case of CDA-DMs, followed by constructing the codes for $L^{\prime}$-PSK to obtain the desired number of DMs. The DMs generated from codes designed for $L^{\prime}$-PSK can be used in conjunction with $L$-PSK, since $L$-PSK is a subset of $L^{\prime}$-PSK, because $L$ and $L^{\prime}$ are powers of 2. For example, one can use DMs derived from FECs and CDA codes constructed for 8-PSK in conjunction with 4-PSK.

It is not known a priori as to which of the available options is optimal for a given rate. Thus, one has to evaluate the performance of each of the options for a given rate for choosing the best configuration. However, the number of options to be evaluated can be shown to be $(RM+1)$, which becomes excessive for systems having a high-rate and a large number of transmit antennas. For example, for a system having a spectral efficiency of 20 bpcu and 4 transmit antennas, 81 STSK configurations have to be evaluated. Furthermore, a configuration optimal for a given rate using CO-DMs is not necessarily optimal for FEC-DMs. Thus, the configuration has to be evaluated in conjunction with the specific DM set used. Furthermore, the complexity involved in the optimization of DMs for coding gain, or capacity will be high for configurations of high $Q$ \cite{DMOPT}. Thus, the computational complexity involved in finding the optimal configuration for high rate systems will be high. However, with the aid of the proposed structured DMs this complexity is significantly reduced as one has to evaluate only the specific configurations available for a given rate. Thus we emphasize that for an STSK scheme using an arbitrary PSK constellation and full-diversity DMs, existing ST codes such as FECs and CDA codes are a convenient choice.

FECs are rate-one codes and hence impose limitations on the achievable rates by restricting the maximum value of $Q$ to $L^{M-1}$, as shown in Section III. Increasing the value of $L$ in order to increase $Q$ obviously affects the coding gain offered, since, the minimum distance of the PSK constellation drops as $|2\sin{(\frac{\pi}{L})}|$, which indirectly determines the achievable coding gain.

Table I shows the drop in coding gain offered by the FEC-DMs and CDA-DMs with the increase in DMs in STSK(2,2,2,$Q$), QAM system. It is clear from the Table I that FEC-DMs are more suitable for low-rate and CDA-DMs are suitable for high-rate applications.

An other class of STBCs subsumed by the class of Decomposable Dispersion Codes is constituted by Linear Constellation Precoding (LCP) based STBCs \cite{GIANNA}. Dispersion Matrices from LCP based codes are attractive alternatives for DMs from FECs, as the former will have a full-rank $\chi$ associated with them. A detailed study of DMs from the LCP based codes is left for our future study. From our study of STBCs and the STSK scheme we conjecture that most of the existing structured linear ST codes are subsumed by the class of DDCs, and hence they are capable of exploiting the low decoding complexity techniques that are specific to the STSK scheme.
\section{Simulation Results and Discussion}

$\mathbf{Simulation~Scenario}$:
In all our simulations we have used at least $10^{t+1}$ symbols, at an SER of $10^{-t}$ for evaluating the SER and assumed a block Rayleigh fading channel. The DMs of {\em Example~1} are used for FEC-DMs and those of {\em Example~2} are used for CDA-DMs with  $t_2=e^{j{\pi \over 2}}$ and $\delta=e^{j{3\pi \over 8}}$ in our simulations. We note here that $t_2$ and $\delta$ are not transcendental over $\mathbb{Q}(S)$, but upon adding a small positive constant $\epsilon$ to the exponentials, they can be made transcendental without significantly affecting the achievable coding gain. For the $Q=8$ case, CO-DMs are generated by maximizing the mutual information over a large set of DMs having complex Gaussian entries satisfying the unity-average power constraint. These DMs are given in Appendix A. For the $Q=2$ and $Q=4 $ cases, the DMs given in \cite{CSTSK} are used.

Table III summarizes the coding gain offered both by the proposed and by the existing schemes for CSTSK(2,2,2,$Q$).

From Table III it is clear that the proposed FEC- and CDA-DMs offer a better coding gain compared to the existing CO-DMs. Hence, the proposed DMs can be expected to give a better SER performance at high SNRs than the CO-DMs. 

Simulation results with perfect and imperfect Channel State Information at the Receiver (CSIR) are presented separately. Under perfect CSIR conditions the performance of the proposed FEC-, CDA-DMs, and of the CO-DMs is evaluated by considering both the ML and MF based detector of \cite{SSLCDET}, for the imperfect CSIR scenario the iterative detection/estimation algorithm of \cite{DETEST} is used with 2 training and 100 data carrying STSK blocks.

\subsection{With Perfect CSIR}

\begin{figure}
\centering
\includegraphics[scale=0.5]{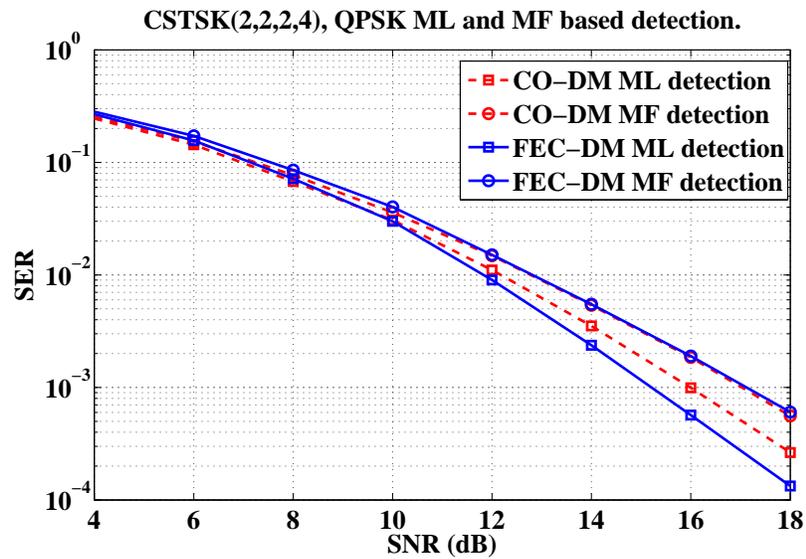}
\caption{SER performance curves of FEC-DMs and CO-DMs in CSTSK(2,2,2,4), QPSK ($R=2$).}
\label{Q4}
\end{figure}

Fig. \ref{Q4} characterizes the SER performance of the FEC-DMs as well as that of the existing CO-DMs in conjunction with the QPSK constellation for the $Q=4$ case. Observe from Fig. \ref{Q4} that the proposed set of DMs gives a better performance than the existing CO-DMs under single-stream based ML detection \cite{LCML}, which is attributed to the higher coding gain of the proposed DM set. It is also clear from the figure that the performance of the FEC-DMs is nearly the same as that of the existing CO-DMs in conjunction with the Matched Filtering (MF) based low-complexity detector \cite{SSLCDET}. Thus, the FEC-DMs suffer from a relatively higher performance loss compared to CO-DMs, when using the MF based detector. This is due to the rank deficiency of the matrix $\chi$ associated with the FEC-DMs. However, the FEC-DM scheme exhibits an SNR gain of about 1dB over the existing scheme at an SER of about $10^{-3}$ for ML detection.

\begin{figure}
\hspace{-30pt}
\centering
\includegraphics[scale=0.5]{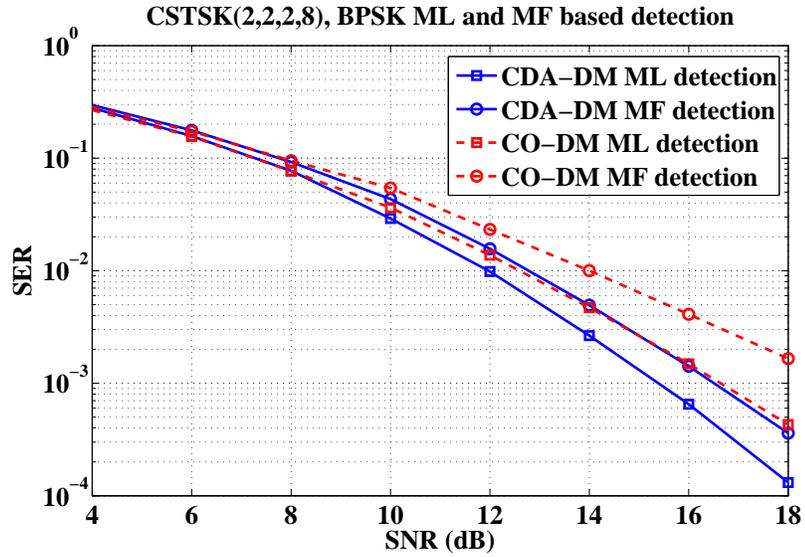}
\caption{SER performance curves of CDA-DM and CO-DM in CSTSK(2,2,2,8), BPSK ($R=2$).}
\label{Q8BPSK}
\end{figure}

Fig. \ref{Q8BPSK} characterizes the SER performance of both the proposed CDA-DMs as well as of the existing CO-DM scheme in conjunction with the BPSK constellation for the $Q=8$ case. We note that the rate of this scheme is the same as that of the scheme using the FEC-DMs, i.e., $R=2$. Due to the better coding gain of the CDA-DMs, the proposed scheme outperforms the CO-DM scheme both for the ML and for the MF based detectors. Unlike FEC-DMs, CDA-DMs have full-rank $\chi$ and hence they do not suffer from any significant performance degradation under MF based detection. We observe that the CDA-DM scheme exhibits an SNR gain of about 1dB over the existing scheme at about an SER of about $10^{-3}$. It is noticeable that the CO-DMs show a slightly higher performance loss compared to their $Q=4$ counterparts, as seen in Fig. \ref{Q4}. This is attributed to the low coding gain and the sensitivity of the MF based detector to a large $Q$. Thus, it is evident from Fig. \ref{Q4} and Fig. \ref{Q8BPSK} that different STSK configurations giving the same rate may perform differently. We also emphasize that the DMs should be optimized for maximum coding gain, rather than for maximum capacity, when aiming for a better BER performance in conjunction with a MF based detector.

\begin{figure}
\hspace{-30pt}
\centering
\includegraphics[scale=0.5]{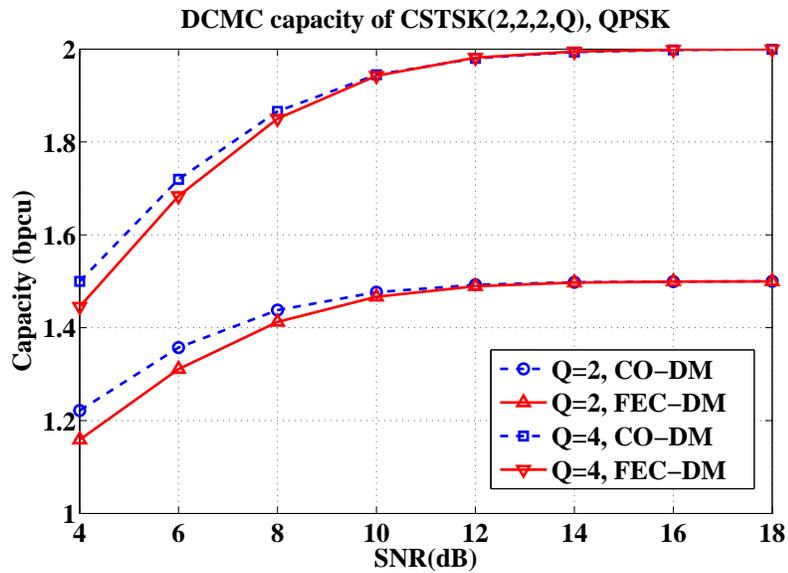}
\caption{DCMC capacity curves of FEC-DMs and CO-DMs in CSTSK(2,2,2,$Q$), BPSK ($R=1.5~\text{and}~2$).}
\label{DCMCQAM}
\end{figure}

\begin{figure}
\hspace{-30pt}
\centering
\includegraphics[scale=0.5]{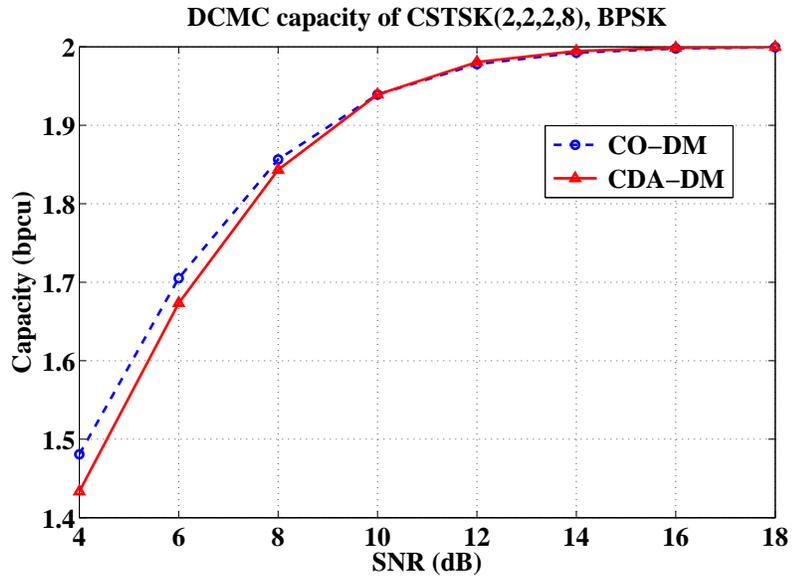}
\caption{DCMC capacity curves of CDA-DMs and CO-DMs in CSTSK(2,2,2,8), BPSK ($R=2$).}
\label{DCMCQ8BPSK}
\end{figure}

Figs. \ref{DCMCQAM} and \ref{DCMCQ8BPSK} show the Discrete-Input Continuous-Output Memoryless Channel (DCMC) capacity of the schemes considered above. More specifically, Fig. \ref{DCMCQAM} shows the DCMC capacity \cite{DCMCREF} curves of CSTSK(2,2,2,$Q$), QPSK for the $Q=2$ and $Q=4$ scenarios. We observe from Fig. \ref{DCMCQAM} that the capacity of the FEC-DM scheme approaches that of the CO-DM scheme for SNRs greater than 10dB for the $Q=4$ case and for SNRs higher than 14dB for the $Q=2$ scenario. Thus, FEC-DMs offer a better coding gain than CO-DMs without capacity loss at high SNRs.

Fig. \ref{DCMCQ8BPSK} shows the DCMC capacity curves of the CSTSK(2,2,2,8), BPSK scheme. We observe from Fig. \ref{DCMCQ8BPSK} that the CDA-DM scheme approaches the capacity as that of the CO-DMs for SNRs greater than 10dB. Thus, CDA-DMs offer a better coding gain than CO-DMs without any capacity loss at medium and high SNRs.

\subsection{With imperfect CSIR}

\begin{figure}
\hspace{-30pt}
\centering
\includegraphics[scale=0.5]{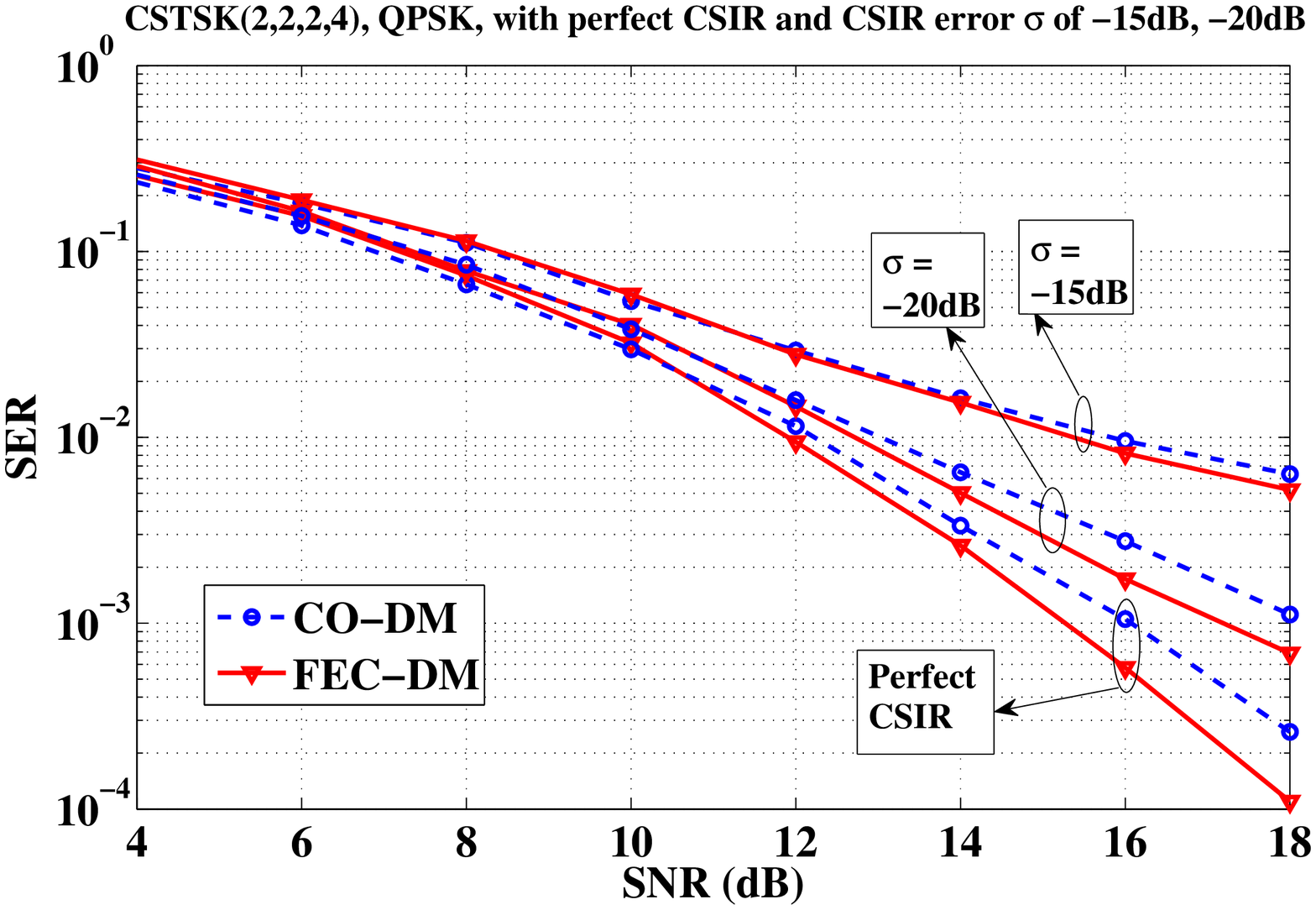}
\caption{Performance of FEC-DMs and CO-DMs in CSTSK(2,2,2,4), QPSK, under perturbed channel conditions with ML detection. }
\label{FECPERT}
\end{figure}

\begin{figure}
\hspace{-30pt}
\centering
\includegraphics[scale=0.5]{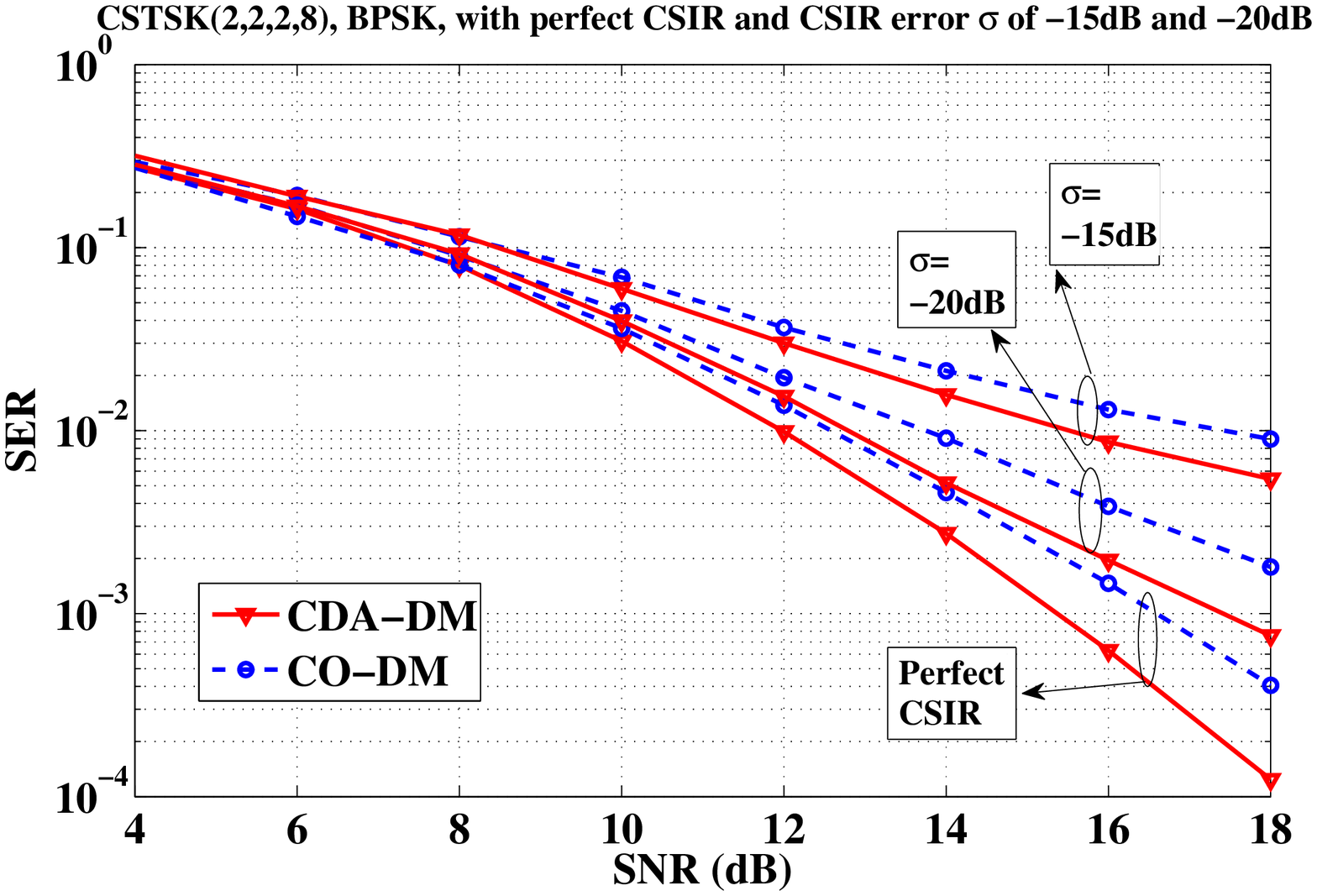}
\caption{Performance of CDA-DMs and CO-DMs in CSTSK(2,2,2,8), BPSK, under perturbed channel conditions with ML detection.}
\label{CDAPERT}
\end{figure}

In this subsection we study the SER performance of both the proposed and of the existing DMs under realistic imperfect CSIR conditions. Fig. \ref{FECPERT} and \ref{CDAPERT} illustrate the sensitivity of the SER performance under ML detection to CSIR perturbations, where $\sigma$ is the variance of the complex-valued circular symmetric Gaussian noise that models the channel estimation error. It is clear from the Figures \ref{FECPERT} and \ref{CDAPERT} that the performance of both the proposed and of the existing DMs degrades upon increasing the channel's estimation error variance. It is evident from the figures that the proposed FEC- and CDA-DMs perform significantly better than the existing CO-DMs for all the estimation error variances considered. This is attributed to their higher coding gains. Thus, we emphasize again that instead of optimizing DMs for capacity or mutual information, DMs should be optimized for coding gain in order to achieve robustness to CSIR perturbations.

\begin{figure}
\hspace{-25pt}
\centering
\includegraphics[scale=0.5]{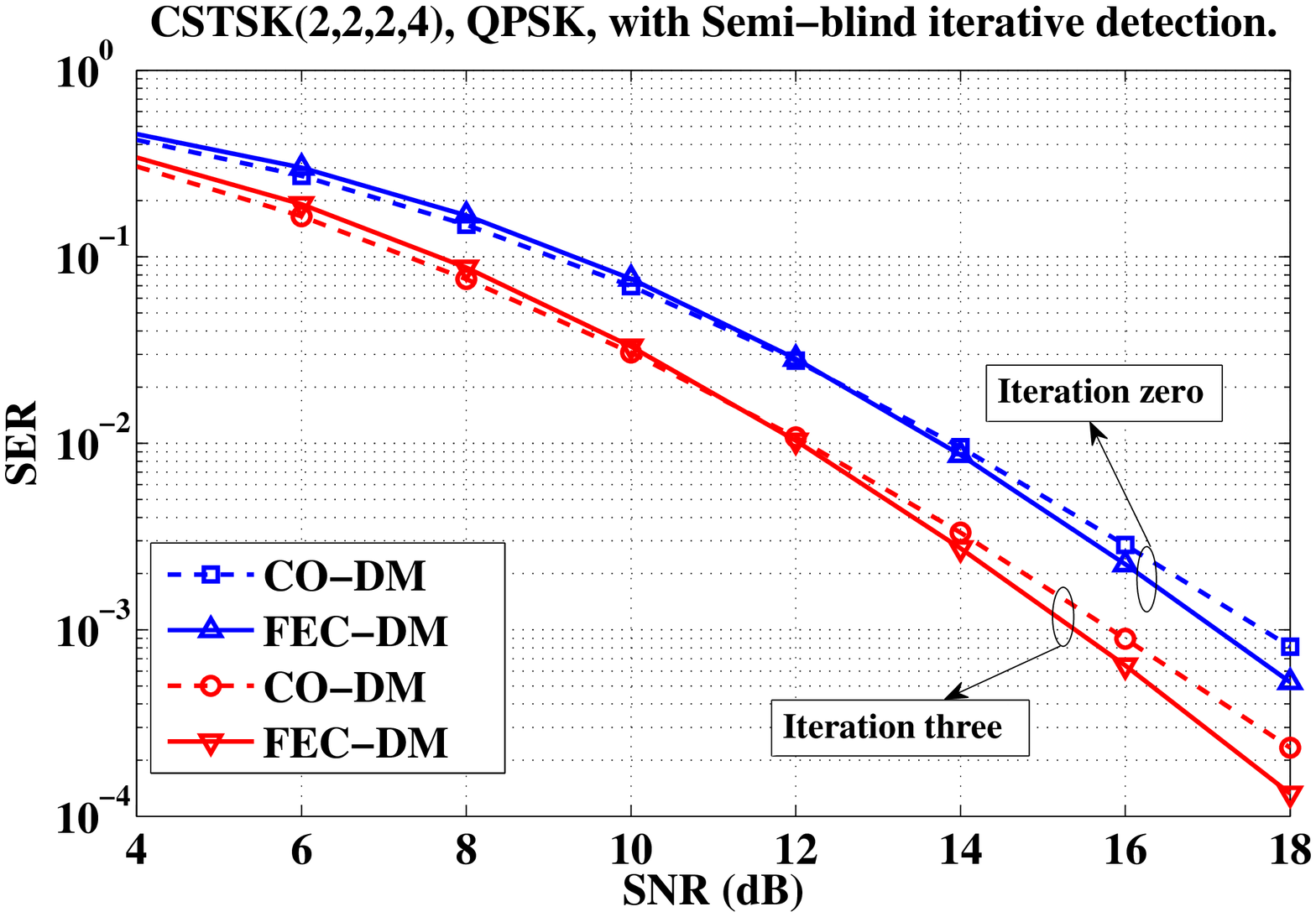}
\caption{Performance of FEC-DMs and CO-DMs with 2 training and 100 data STSK blocks with semi-blind iterative detection/estimation algorithm at the receiver. }
\label{FECITERDET}
\end{figure}

\begin{figure}
\hspace{-25pt}
\centering
\includegraphics[scale=0.5]{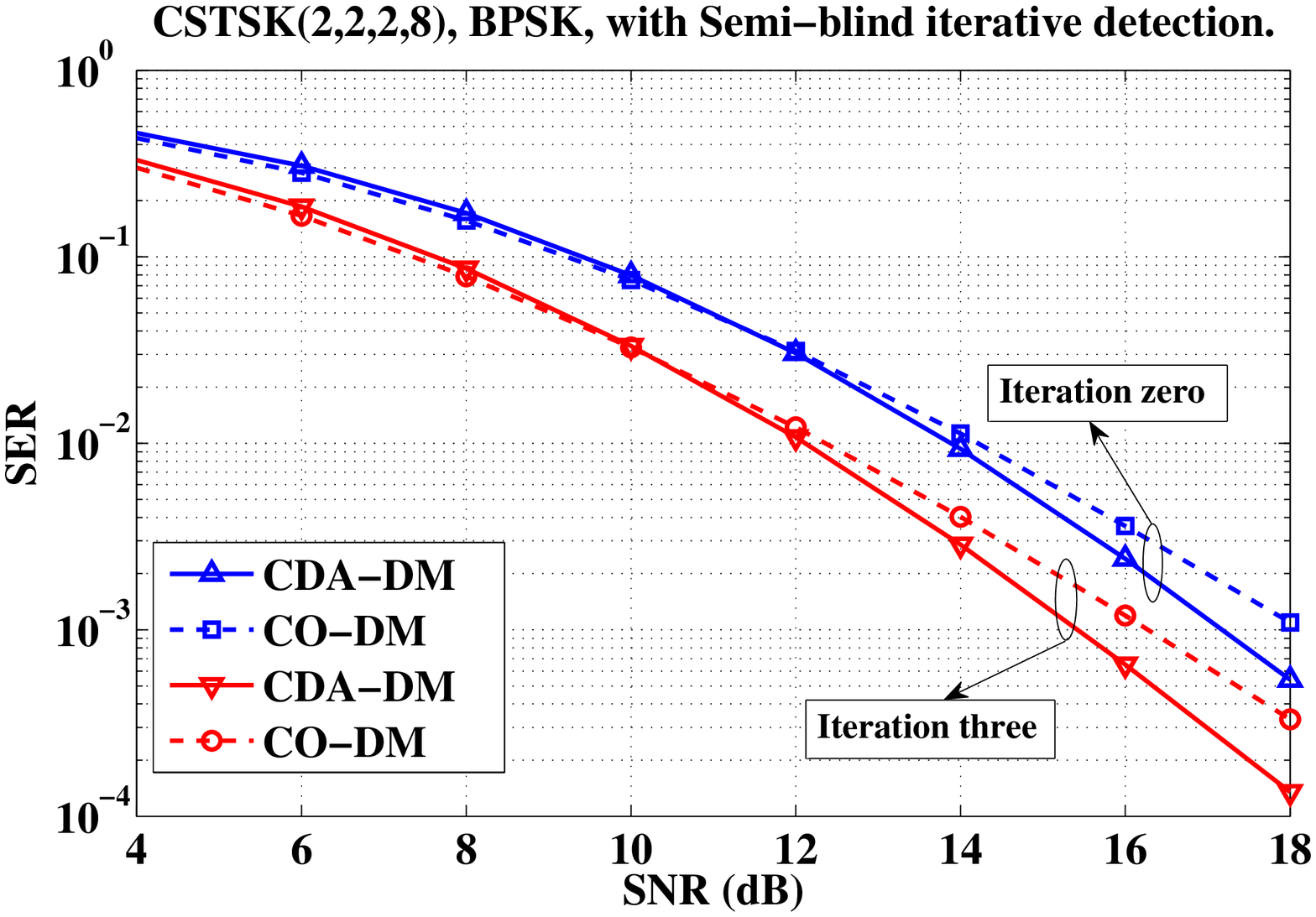}
\caption{Performance of FEC-DMs and CO-DMs with 2 training and 100 data STSK blocks with semi-blind iterative detection/estimation algorithm at the receiver.}
\label{CDAITERDET}
\end{figure}

A semi-blind iterative detection/estimation algorithm was proposed for STSK systems in \cite{DETEST} which uses an initial Least Squares channel estimate and then iteratively detects the transmitted ST matrices and estimates the channel with the aid of the detected data. Fig. \ref{FECITERDET} and Fig. \ref{CDAITERDET} show the SER performance of both FEC-DMs and CDA-DMs against their CO-DMs counterparts respectively, with the above-mentioned detection/estimation algorithm based receiver. It was observed from our simulations that there is no significant performance gain beyond the third iteration in the above mentioned algorithm. Thus, the SER curves of both the proposed and of the existing DMs are presented for iterations zero and  three only. It is evident from the figures that the proposed DMs give a better SER performance than the CO-DMs for SNRs higher than 12dB. In CSTSK(2,2,2,8), BPSK, the proposed CDA-DMs have shown an SNR improvement of 1dB at an SER of about $10^{-3}$ with respect to the CO-DMs. By comparing the SER curves corresponding to the perfect CSIR condition of Figs.  (\ref{FECPERT}, \ref{CDAPERT}) to the curves corresponding to iteration three of Figs. (\ref{FECITERDET}, \ref{CDAITERDET}), it is evident that the proposed DMs approach ML performance.

\section{Conclusions}

We demonstrated in this paper that the low-complexity decoding benefits of the STSK scheme are not specific to the STSK family, but are applicable to most of the structured STBCs constructed for arbitrary PSK signal sets. A class of STBCs that falls into the STSK framework has been identified and termed as Decomposable Dispersion Codes. It is shown that the codes from this class also enjoy the low-complexity decoding benefits that are specific to the STSK scheme. Both FECs and CDA codes were shown to belong to this class, which resulted in DMs with beneficial coding gains without compromising on the achievable capacity. The proposed DMs have attained a better SER performance due to their higher coding gain under both perfect and imperfect CSIR conditions both with ML and with matched filtering based detectors. It is also observed from our simulation results that using the high coding gain DMs is imperative for achieving robustness to channel estimation errors.

\section{Appendix A}
$\bullet$ The DMs optimized for CSTSK(2,2,2,8), BPSK constellation obtained by maximizing the mutual information are:
\[
 \mathbf{A}_1= \left[\begin{array}{cc} -0.2609 - j0.1663 &    0.4274 + j1.2471\\ -0.3356 - j0.1604 & \text{~~}0.0127 + j0.1667\\\end{array}\right],
 \]
\[
 \mathbf{A}_2=\left[\begin{array}{cc} -0.8256 + j0.5391 &    0.1502 + j0.0534\\  -0.0718 - j0.4744 &  \text{~~}  0.3378 - j0.8112\\\end{array}\right],
 \]
\[
 \mathbf{A}_3=\left[\begin{array}{cc} -0.4371 - j0.3679 &   -0.5509 - j0.3024\\  -0.8711 + j0.1085 &   -0.4850 - j0.5224\\\end{array}\right],
 \]
\[
 \mathbf{A}_4=\left[\begin{array}{cc} -0.1173 - j0.8969 &     0.1467 + j0.2945\\   -0.2049 + j0.4875 &  \text{~~}  0.8546 + j0.2524\\\end{array}\right],
 \]
\[
 \mathbf{A}_5= \left[\begin{array}{cc} -0.0852 - j0.1935 &    0.6287 + j0.0950\\  0.9992 - j0.3717 &     -0.5449 - j0.3428\\\end{array}\right],
 \]
\[
 \mathbf{A}_6=\left[\begin{array}{cc} -0.2352 +j1.0560 &    -0.6267 - j0.1166\\  0.1142 + j0.4872 &    -0.4154 + j0.0112\\\end{array}\right],
 \]
\[
 \mathbf{A}_7=\left[\begin{array}{cc} -0.1408 + j0.0534 &     -0.4832 + j0.8613\\    0.6937 + j0.6212 &    0.1325 - j0.3425\\\end{array}\right],
 \]
\[
 \mathbf{A}_8=\left[\begin{array}{cc} -0.4118 + j0.0950 &     0.6746 - j0.0363\\  -0.5485 + j0.3372 &     -0.9707 + j0.0908\\\end{array}\right].
 \]

\section{Appendix B\\Proof of Theorem 1}

{\em Lemma ~B1:} Any {\em unrotated} PSK signal set with cardinality $L$ is a cyclic group under multiplication with group generator $g=e^{j\frac{2\pi}{L}}$ and identity element $i_g=1$.

{\em Proof:} Straightforward.

\begin{proof}[\hspace{-10 pt}Proof of Theorem 1]
Let the set $(S\times {\cal E})$ be denoted by ${\cal E}^{\prime}$. An arbitrary element of ${\cal E}^{\prime}$ is given by $ \Big(f_i, {\mathbf{M}_l}+\sum_{i=0,i\neq l}^{V-1}f_i^\prime{\mathbf{M}_i} \Big), ~\text{where} ~f_i,f_i^\prime \in S$. Since, $\zeta_p$ is a product mapping it maps this element to $f_i{\mathbf{M}_l}+\sum_{i=0,i\neq l}^{V-1}f_if_i^\prime{\mathbf{M}_i}$. From {\em Lemma~B1}, we can write this element as $\Big(g^q{\mathbf{M}_l}+\sum_{i=0,i\neq l}^{V-1}g^{{(q+q_i^\prime)}_L}{\mathbf{M}_i}\Big)$,
where $f_i=g^q$, $f_i^\prime=g^{q_i^\prime}$ for some $0 \leq q,\{q_i^\prime\}_{i=0,i\neq l}^{V-1} \leq L-1$, and ${(q+q_i^\prime)}_L=({q+q_i^\prime}) \mod L$. Letting ${(q+q_i^\prime)}_L=q_i^{\prime\prime}$ we have, $ \Big(g^q{\mathbf{M}_l}+\sum_{i=0,i\neq l}^{V-1}g^{{q_i^{\prime\prime}}}{\mathbf{M}_i}\Big)~\in {\cal C}$,
since, $0\leq \{q_i^{\prime\prime}\}_{i=0,i\neq l}^{V-1}\leq L-1$. Thus, we have shown that $\zeta_p$ is a mapping from $(S\times {\cal E})$ to ${\cal C}$.
Now, we proceed to prove that it is a bijection.
The cardinality of the set $S$ is $L$ and by careful inspection of Eq.(\ref{ELEQDDC}) (Eq.(\ref{KLEQ}) for {\em Theorem~2}), we have
$|{\cal E}|=L^{V-1}$. Thus, we have $|S\times {\cal E}|=L\cdot L^{V-1}=L^V.$ From Eq.(\ref{LDCEQN}) (Eq.(\ref{KEQ}) for {\em Theorem~2}) we infer that $|{\cal C}|=L^V$, when $F=S$.
Therefore, $|S\times {\cal E}|=|{\cal C}|=L^V$. Hence, to prove that $\zeta_p$ is a bijection, it is sufficient to prove that it is a one-to-one mapping.
Any two elements of ${\cal E}$ may be written as:
\begin{equation*}
k_{r,\{r_i\}}^\prime= \Big(g^r,{\mathbf{M}_l}+\sum_{i=0,i\neq l}^{V-1}g^{r_i}{\mathbf{M}_i}\Big), 
\end{equation*}
\begin{equation}
k_{q,\{q_i\}}^\prime= \Big(g^q,{\mathbf{M}_l}+\sum_{i=0,i\neq l}^{V-1}g^{q_i}{\mathbf{M}_i}\Big),
\end{equation}
for some $0\leq r,q,\{r_i,q_i\}_{i=0,i\neq l}^{V-1}\leq L-1$. The above elements are distinct if $r \neq q$ or $\{r_i \neq q_i\}$
for any $0\leq i\leq V-1$ such that $i\neq l$. When $r \neq q$, 
\begin{equation*}
\zeta_p:k_{r,\{r_i\}}^\prime \mapsto k_{r,\{r_i\}}=\Big(g^r{\mathbf{M}_l}+\sum_{i=0,i\neq l}^{V-1}g^{(r_i+r)_L}{\mathbf{M}_i}\Big), 
\end{equation*}
\begin{equation}
\zeta_p:k_{q,\{q_i\}}^\prime \mapsto k_{q,\{q_i\}}=\Big(g^q{\mathbf{M}_l}+\sum_{i=0,i\neq l}^{V-1}g^{(q_i+q)_L}{\mathbf{M}_i}\Big),
\end{equation}
it is straightforward that $k_{r,\{r_i\}}\neq k_{q,\{q_i\}}$. When $r=q$ and $\{r_i \neq q_i\}$ for any $0\leq i\leq V-1$ such that $i\neq l$,
we have $k_{r,\{r_i\}}\neq k_{q,\{q_i\}}$, since, $\exists$ at least one $i=i^\prime$ for which $r_i\neq q_i$, and hence
$(r+r_i)_L\neq (r+q_i)_L$. Thus, $\zeta_p$ is a one-to-one mapping as it maps distinct elements in its domain to distinct elements of its co-domain.
\end{proof}

\section{Appendix C\\ Proof of Theorem 3}
\begin{proof}
From {\em Lemma~B1} and with $F=S$ in Eq.(\ref{KEQ}), we can write any arbitrary element in $K$ as
\begin{equation} 
 \sum_{i=0}^{M-1}g^{q_i}(t_M)^i, ~0\leq\{q_i\}_{i=0}^{M-1}\leq L-1.
\end{equation}
Consider an arbitrary element in $S$ given by $g^q$ for some $0\leq q\leq L-1$. Thus, $\zeta_p$
maps the element $(g^q,\sum_{i=0}^{M-1}g^{q_i}(t_M)^i)$ to $\sum_{i=0}^{M-1}g^{q_i^{\prime\prime}}(t_M)^i$,
where $q_i^{\prime\prime} =(q+q_i)_L$. It is straightforward to show that for each $i$, $q_i^{\prime\prime}$ spans the set $\{0,1,\hdots,L-1\}$ as $q_i$ varies from $0$ to $L-1$ for any $q$. Thus, $\zeta_p$ maps $~S\times K$ to $K$ and $\zeta_p: (g^q, K)\mapsto K$ is a one-to-one mapping for any given $g^q\in S$.
\end{proof}

\section{Appendix D\\ Proof of Theorem 4}
\begin{proof}
 From the proof of {\em Theorem~1} it is straightforward to show that $\zeta_p$ is a mapping from $S_{sym}\times {\cal E}$ to $\cal C$. Since, $|{\cal E}|=|S^{\prime}|\cdot L^{V-1}=(L/{|S_{sym}|})\cdot L^{V-1}$, we have $|S_{sym}\times {\cal E}|=|S_{sym}|\cdot (L/{|S_{sym}|})\cdot L^{V-1}=L^V=|{\cal C}|$. Thus, the domain and co-domain of $\zeta_p$ have the same number of elements. Hence, to show that $\zeta_p$ is a bijection, it is sufficient to show that it is a one-to-one mapping. Any two elements of $S_{sym}\times{\cal E}$ can be formulated as:
\begin{equation*}
 \Big(g^q,s_kg^{m_k}{\mathbf{M}_l}+\sum_{i=0,i\neq l}^{V-1}s_{k_i}g^{n_i}{\mathbf{M}_i}\Big), 
\end{equation*}
\begin{equation}
\Big(g^r,s_pg^{m_p}{\mathbf{M}_l}+\sum_{i=0,i\neq l}^{V-1}s_{p_i}g^{t_i}{\mathbf{M}_i}\Big),
\end{equation}
where $0\leq q,r, \{n_i,t_i\}_{i=0,i\neq l}^{V-1} \leq |S_{sym}|-1$, $1\leq k,p,\{k_i,p_i\}_{i=0,i\neq l}^{V-1} \leq L/4$ . The map $\zeta_p$ maps these elements to 
\begin{equation*}
 \Big(s_kg^{m_k+q}{\mathbf{M}_l}+\sum_{i=0,i\neq l}^{V-1}s_{k_i}g^{n_i+q}{\mathbf{M}_i}\Big),
\end{equation*}
\begin{equation*}
 \Big(s_pg^{m_p+r}{\mathbf{M}_l}+\sum_{i=0,i\neq l}^{V-1}s_{p_i}g^{t_i+r}{\mathbf{M}_i}\Big),
\end{equation*}
respectively. For $k \neq p$ the above terms are unequal, since $s_k$ and $s_p$ correspond to different points in $S_{L/4}$. For $k=p$ and $q\neq r$, the above terms are unequal, since $g^{m_k+q} \neq g^{m_k+r}$. For $k=p$, $q=r$, $\exists$ at least one $i$ in $\{i\}_{i=0,i\neq l}^{V-1}$ such that $k_i\neq p_i$ or $n_i \neq t_i$ since the two elements are distinct. The above terms are unequal if $k_i\neq p_i$, since $s_{k_i}$ and $s_{p_i}$ correspond to different elements in $S_{L/4}$. For $k_i=p_i$ and $n_i \neq t_i$ it is straightforward to show that $g^{n_i+q}\neq g^{t_i+q}$ and hence the above terms are unequal. Thus, $\zeta_p$ maps distinct elements in its domain to distinct elements in its co-domain, and hence is a one-to-one mapping.

\end{proof}

\bibliographystyle{ieeetr}

\begin{table}
\caption{Coding gain offered by the proposed DMs in CSTSK(2,2,2,$Q$), QAM.}
\begin{center}
\begin{tabular}{|c||c|c|c|}\hline
	\small No. of DMs &  \small $Q=4$ & \small $Q=16$& \small $Q=64$\\
	\hline\hline
  \small $G$ (FEC-DM) &  \small 1 &  \small 0.0058 & \small 0.00002318\\
	\hline 
\small $G$ (CDA-DM) &  \small 0.25 &   \small 0.1464 & \small 0.0565\\
	\hline 
\small $t_2$ & \small $e^{\frac{j3\pi}{16}}$ & \small $e^{\frac{j3\pi}{8}}$ & \small $e^{\frac{j\pi}{4}}$\\
\hline
\small $\delta$ & \small $e^{\frac{j\pi}{2}}$ & \small $e^{\frac{j3\pi}{4}}$ & \small $e^{\frac{j27\pi}{16}}$\\
	\hline
\end{tabular}\\
\end{center}
\end{table}

\begin{table}
\caption{Various possibilities of $Q$ and $L$ in CSTSK(2,2,2,$Q$) for $R$ bpcu.}
\begin{center}
\begin{tabular}{|c||c|c|c|c|c|c|}\hline
 \small $R$ bpcu  &  \small Option 1 & \small Option 2 & \small Option 3 & \small Option 4 & \small Option 5 & \small Option 6 \\
\hline
\hline
\small 1 & \small $Q=1$, $L=4$ & \small $Q=2$, $L=2$ & \small $Q=4$, $L=1$ & \small * & \small * & \small *\\
\hline
\small 1.5 & \small $Q=1$, $L=8$  & \small $Q=2$, $L=4$ & \small $Q=4$, $L=2$  & \small $Q=8$, $L=1$ & \small * & \small * \\
\hline
\small 2 & \small $Q=1$, $L=16$ & \small $Q=2$, $L=8$ & \small $Q=4$, $L=4$ & \small $Q=8$, $L=2$ & \small  $Q=16$, $L=1$ & \small *\\
	\hline
\small 2.5 & \small $Q=1$, $L=32$ & \small $Q=2$, $L=16$ & \small $Q=4$, $L=8$ & \small $Q=8$, $L=4$ & \small $Q=16$, $L=2$ & \small $Q=32$, $L=1$ \\
	\hline
\end{tabular}\\
\end{center}
\end{table}

\begin{table}
\caption{Comparison of coding gains offered by the example constructions.}
\begin{center}
\begin{tabular}{|c||c|c|}
	\hline
 \small &  \small $Q=4$ & \small $Q=8$\\
\small   Schemes & \small \hspace{10pt} QPSK \hspace{10 pt} & \small \hspace{2pt} BPSK \hspace{10 pt}\\
	\hline\hline
 \small CO-DM &  0.1882 & 0.0455\\
	\hline
 \small FEC-DM &  1 & *\\
\hline
\small CDA-DM  &  * & 1\\
	\hline 
\end{tabular}\\
\end{center}
\end{table}

\end{document}